\documentclass[10pt,journal,compsoc]{IEEEtran}

\usepackage{ifpdf}        %
\usepackage{amsmath}      %
\usepackage{algorithmic}  %
\usepackage{array}        %
\usepackage{url}          %

\ifpdf
\else
\fi

\ifCLASSOPTIONcompsoc
  \usepackage[nocompress]{cite}
\else
  \usepackage{cite}
\fi

\ifCLASSINFOpdf
   \usepackage[pdftex]{graphicx}
   \graphicspath{{figures/}{pictures/}{images/}{./}}  %
   \DeclareGraphicsExtensions{.pdf,.png,.jpg,.jpeg}   %
\else
   \usepackage[dvips]{graphicx}
   \graphicspath{{figures/}{pictures/}{images/}{./}}  %
   \DeclareGraphicsExtensions{.eps}                   %
\fi

\ifCLASSOPTIONcompsoc
 \usepackage[caption=false,font=footnotesize,labelfont=sf,textfont=sf]{subfig}
\else
 \usepackage[caption=false,font=footnotesize]{subfig}
\fi

\ifCLASSOPTIONcaptionsoff
 \usepackage[nomarkers]{endfloat}
\let\MYoriglatexcaption\caption
\renewcommand{\caption}[2][\relax]{\MYoriglatexcaption[#2]{#2}}
\fi

\usepackage{enumitem}            %
\usepackage[dvipsnames]{xcolor}  %
\usepackage[normalem]{ulem}      %
\usepackage{hyperref}            %
\usepackage[letterspace=-100]{microtype}

\definecolor{mred}{rgb}{.80,.12,.30}

\newif\ifauthornotes
\newif\ifstrike
\newif\ifadd
\newif\ifreplace
\newif\iftodo

\newcommand{\strike}[1]{\ifstrike{\color{mred}{\texorpdfstring{\sout{#1}}{#1}}}\fi}

\newcommand{\add}[1]{\ifadd{\leavevmode\color{magenta}{#1}}\else{#1}\fi}

\newcommand{\iitfc}[0]{{\fontfamily{qcr}\selectfont\scshape\lsstyle Combined}}
\newcommand{\niitfc}[0]{{\fontfamily{qcr}\selectfont\scshape\lsstyle Separated}}

\begin{document}

\title{Preliminary Guidelines For Combining Data Integration and Visual Data Analysis}

\author{
  Adam~Coscia,
  Ashley~Suh,
  Remco~Chang,
  and~Alex~Endert%
  \IEEEcompsocitemizethanks{
    \IEEEcompsocthanksitem Adam Coscia and Alex Endert are with Georgia Institute of Technology. Emails: \{acoscia6, endert\}@gatech.edu.
    \IEEEcompsocthanksitem Ashley Suh and Remco Chang are with Tufts University. Emails: ashley.suh@tufts.edu, remco@cs.tufts.edu.
  }%
  \thanks{\textcolor{red}{This manuscript is a Preprint Version - Accepted, to Appear: IEEE Transactions on Visualization and Computer Graphics, DOI: 10.1109/TVCG.2023.3334513}}
}%

\IEEEtitleabstractindextext{%
  \begin{abstract}
    Data integration is often performed to consolidate information from multiple disparate data sources during visual data analysis.
    However, integration operations are usually separate from visual analytics operations such as encode and filter in both interface design and empirical research.
    We conducted a preliminary user study to investigate whether and how data integration should be incorporated directly into the visual analytics process.
    We used two interface alternatives featuring contrasting approaches to the data preparation and analysis workflow: manual file-based ex-situ integration as a separate step from visual analytics operations; and automatic UI-based in-situ integration merged with visual analytics operations.
    Participants were asked to complete specific and free-form tasks with each interface, browsing for patterns, generating insights, and summarizing relationships between attributes distributed across multiple files.
    Analyzing participants' interactions and feedback, we found both task completion time and total interactions to be similar across interfaces and tasks, as well as unique integration strategies between interfaces and emergent behaviors related to satisficing and cognitive bias.
    Participants' time spent and \add{interactions}\strike{ emergent strategies} revealed that in-situ integration enabled users to spend more time on analysis tasks compared with ex-situ integration.
    \add{Participants' integration strategies and analytical behaviors revealed differences in interface usage for generating and tracking hypotheses and insights}\strike{, yet their emergent behaviors suggested that in-situ integration could negatively affect the ability to generate and track hypotheses and insights}.
    With these results, we synthesized preliminary guidelines for designing future visual analytics interfaces that can support integrating attributes throughout an active analysis process.
  \end{abstract}

}

\maketitle

\IEEEdisplaynontitleabstractindextext

\IEEEraisesectionheading{
  \section{Introduction}
  \label{sec:introduction}
}

\IEEEPARstart{F}{}rom a visual analytics perspective, the rapid growth of data today requires methods to combine information from disparate sources into a unified data representation to facilitate analytical reasoning \cite{Thomas:2006:VisualAnalyticsAgenda, Keim:2008:VADefinitionProcessChallenges}.
From a systems engineering perspective, this process involves data integration, or the task of querying multiple, often heterogeneous, data sources with potentially differing levels of access and resolving the results into a unified view of the data \cite{Lenzerini:2002:TheoryDataIntegration, Doan:2012:PrinciplesDataIntegration}.
A human-in-the-loop perspective promotes ``exploratory knowledge discovery in large datasets'' where \textit{a priori} knowledge of data is not guaranteed \cite{Endert:2014:HumanIsTheLoop}.
Yet visual analytics tools such as Tableau present manual data preparation solutions that occur as a separate step from visual analytics operations such as encode and filter \cite{Yi:2007:InteractionTaxonomy}.
We posit that the data preparation and visual analytics workflow in tools like Tableau has created an expectation in research and design that users' interactions and behaviors are influenced by the integration or analysis process separately.

In response, we raise two open research questions based on the common approach of separating data integration and visual analytics processes in research and design.

\begin{enumerate}[topsep=4pt, itemsep=4pt]
    \item \add{\textbf{Where and how should data integration operations, such as joins, be supported in tandem with visual analytics operations, such as encode and filter?}}
\end{enumerate}

\noindent
Kandel et al. identify breakdowns in analysis workflows that occur in the early stages and when transitioning between tasks, where little research and few tools provide visualization solutions \cite{Kandel:2012:Enterprise}.
Theories of information foraging \cite{Pirolli:1999:InformationForaging} and sensemaking processes \cite{Pirolli:2005:SensemakingProcess, Klein:2007:DataframeSensemaking} that describe the analysis process also maintain a simultaneous and inseparable view of continuously finding and making sense of data.
Consider decision-makers integrating columns and rows across multiple spreadsheets using Microsoft Excel to visualize the results as they work.
Dimara and Stasko assert there is a gap in visualization tools that support in-situ data integration around maintaining flexibility and flow when performing visual analytics operations \cite{Dimara:2021:DecisionMakingTasks}.

\begin{enumerate}[topsep=4pt, itemsep=4pt]
    \setcounter{enumi}{1}
    \item \add{\textbf{How will incorporating data integration into an on-going visual analytics process affect user behaviors?}}
\end{enumerate}

\noindent
Pirolli and Card identify potential constraints on analysis during foraging and sensemaking due to time pressures, data overload, and cognitive biases \cite{Pirolli:2005:SensemakingProcess}.
These constraints may lead to satisficing based on the time and effort people spend finding, gathering, and integrating data \cite{Heuer:1999:IntelligenceAnalysis}.

Our aim in this paper is to contribute preliminary guidelines for incorporating data integration into an active visual analytics process, towards fostering better information retrieval that allows people to incorporate their data seamlessly and improve how visualizations are created and used.
To do this, we created two interface alternatives inspired by Polestar \cite{Wongsuphasawat:2015:Voyager} and ran a two (interfaces) by two (data sets) within-subjects study, recruiting 16 participants and randomly assigning an order to use both interfaces and data sets for performing specific and free-form style visual data analysis tasks.
The first interface ({\niitfc}) requires manual file-based ex-situ integration of attributes via Microsoft Excel (column concatenation then selection).
This more traditional interface represents a simplified data preparation and analysis workflow common in most research and design that can serve as a baseline for gathering users' analytical behaviors.
The second interface ({\iitfc}) presents automatic user-interface-based (UI-based) in-situ integration combined with common visual analytics operations such as encode and filter (column concatenation via selection).
This non-traditional interface removes much of the separation between data preparation and visual analytics operations to investigate how users' strategies and analytical behaviors differ from using the first interface.
We then conducted a mixed-methods analysis of users' interactions and behaviors.
As a first step in tackling our broad research questions in a productive way, our approach: (1) reduced confounds in relating users' interactions and behaviors between interfaces; (2) helped us elicit rich qualitative insights that raise important questions and lay the foundation for future work; and (3) mirrored real decision-making scenarios, e.g., Dimara and Stasko \cite{Dimara:2021:DecisionMakingTasks}.

\strike{Comparing participants' analysis processes across interfaces, }\add{We found participants using} unique in-situ integration strategies (Sect.~\ref{sec:integration_strategies}) based on\strike{ the distribution of} time spent integrating, interactions with different panels, and qualitative feedback.
For example, several participants exclusively integrated on the fly on purpose, spending little to no time integrating beforehand.
Yet surprisingly, we found that interface and task type did not significantly affect overall task completion time (Sect.~\ref{sec:time_spent}) or the total number of interactions (Sect.~\ref{sec:interactions}).
At the same time, in-situ integration operations sometimes negatively affected the ability to generate and track hypotheses and insights; specifically, participants' analytical behaviors underscored issues of satisficing and exhibiting biased behaviors (Sect.~\ref{sec:analytic_behaviors}).
With these findings, we synthesized preliminary guidelines for incorporating data integration into visual data analysis: (1) show where and how data are being integrated (Sect.~\ref{sec:guideline_show_data}); (2) use in-situ integration for exploring the space of attributes (Sect.~\ref{sec:guideline_when_insitu}); and (3) balance manual and automated approaches (Sect.~\ref{sec:guideline_balance_approaches}).
We also discuss limitations and future work on eliciting effects in ``real-world'' data integration and visual data analysis scenarios that can involve extensive planning and custom tool development (Sect.~\ref{sec:limitations_future}).

In summary, our work contributes: (1) a within-subjects user study employing two different visual analytics interfaces for integrating and visualizing attributes across multiple disparate data sources; (2) observations and reflections on user interactions and behaviors with our combined data integration and visual analytics interfaces; and (3) preliminary guidelines for incorporating data integration into visual analytics processes and directions for future work.

\section{Related Work}
\label{sec:related_work}

\subsection{Data Integration}
\label{sec:data_integration}
Issues in data integration include specifying well-structured queries, scaling with the number of sources, resolving the heterogeneity of different file formats and data types, and addressing the privacy and accessibility of each source \cite{Doan:2012:PrinciplesDataIntegration}.
Both enterprise and ad-hoc analysis, traditionally utilizing structured data queried from data warehouses with organized schematic definitions, are seeking to incorporate valuable unstructured and semi-structured sources such as PDFs, news feeds, social media, images and video \cite{Morton:2012:TableauDataBlending, Kandel:2012:Enterprise, VanKleek:2013:CarpeData, Hendler:2014:DataIntegrationHeterogenous}.
Further, data sources are increasingly being made publicly available on the Web from government and non-profit organizations such as data.gov and WikiMedia, a movement known as open data \cite{Bizer:2009:WebLinkedData}.
As the scale, availability, and complexity of data sources grows, visual analytics tools should explore ways to incorporate and support the integration process \add{throughout analysis}\strike{ more seamlessly}.

However, locating, collecting and integrating data sources remains an open challenge.
Heterogeneity between data sets, insufficient semantics to describe data, and errors in data insertion and modification \cite{Gal:2014:UncertainEntity} create entity resolution challenges.
For example, when labeled data items with identical attributes from different sources could refer to the same real-world entity, deduplication is needed to identify and link those sets of records.
Further, there are considerable technical challenges in detecting, linking, removing, and merging entities efficiently \cite{Konda:2016:Magellan}.
Work in natural language processing (NLP) has studied how entity resolution can be learned and improved over time through user interactions \cite{Li:2020:DeepEntityMatching}.
A number of tools \cite{Hoefler:2014:LinkedDataWizard, Mohamed:2022:RDFFrames} help people model queries into knowledge graphs \cite{Bonatti:2019:Knowledge, Hogan:2021:KnowledgeGraphs}; however, there are scale limitations for large, complex, and heterogeneous structures \cite{Lou:2019:Knowledge}.
Other automated approaches feature interactive programming interfaces \cite{Dallachiesa:2013:NADEEF}, equi-join-able tables \cite{Zhu:2017:AutoJoin}, and deep learning approaches to entity resolution \cite{Mudgal:2018:Deep}.
As data integration solutions mature, we seek a baseline understanding of user interactions and behaviors in visual analytics tools that incorporate data integration capabilities.

\subsection{Visual Analytics}
\label{sec:visual_analytics}
In-situ data transformations in visual analytics tools often seek to resolve issues of heterogeneity, quality, and semantics, i.e., data wrangling \cite{Kandel:2011:DirectionsDataWrangling, Kandel:2011:Wrangler}.
In this study, we focus on a direct manipulation \cite{Hoefler:2014:LinkedDataWizard, Kahng:2016:ETable} approach to transformations that combine data from separate sources into a single repository, i.e., data integration.
From this perspective, we discuss relevant systems and domains that use interactive interfaces to integrate data throughout the visual analytics process.

\medskip
\noindent\textbf{Systems. }
A few visual analytics systems and studies have addressed the technical and cognitive limitations of integrating data throughout the analysis process in unique ways.
Tableau provides data blending, a technique for dynamically combining data from multiple heterogeneous data sources without any upfront integration effort \cite{Morton:2012:TableauDataBlending}.
Cramer et al. investigate the effects of streaming new data during analysis on sensemaking capabilities, finding an increase in people's explicit focus and reflection on analytic progress \cite{Cramer:2017:StreamingDataSensemaking}.
Cashman et al.'s CAVA system allows users to interactively augment related attributes from knowledge graphs into an existing data set and visualize them during exploration and analysis tasks \cite{Cashman:2020:CAVA}.
Similarly, Latif et al. utilize EventKG in a visualization system to automatically integrate relevant event information and relationships for historical figures in an existing data set \cite{Latif:2021:VisuallyConnecting}.
A goal that cuts across these examples is to help people explore differently structured data across independently located sources without interrupting the visual analytics process \cite{Smith:2006:Facetmap, Bernard:2014:Visual}.
Our aim is to empirically describe user interactions and behaviors when coupling both data integration and visual analytics operations (UI) and processes (workflows).

\medskip
\noindent\textbf{Domains. }
Data integration challenges can be found embedded in various communities that engage in visual data analysis.
Kandel et al. identify data integration as a challenge for analysts in analytics, biology, datamart, finance, healthcare, insurance, marketing, media, retail, social networking, sports, and web development \cite{Kandel:2012:Enterprise}.
Zheng et al. find that urban computing studies routinely integrate and visualize traffic data for public safety and security applications from large, open data sources that are often unstructured, noisy, and heterogeneous \cite{Zheng:2014:Urban}.
In recent work by Dimara and Stasko \cite{Dimara:2021:DecisionMakingTasks}, recent visualization software is criticized by decision-makers for its inability to restructure, integrate, and forage for new data on the fly.
Instead, flexible data software like Excel is preferred by decision-makers, as data is often unstructured and incomplete before analysis begins.
Adding new attributes to an otherwise fixed data set towards improving model performance is an active area of research in machine learning, commonly referred to as data augmentation \cite{Kanter:2015:DeepFeature}.
These challenges have inspired us to ask how the data integration process affects the visual analytics process commonly used across these domains.

\section{Study Design}
\label{sec:study}

We conducted a user study to empirically describe user interactions and behaviors when data integration and visual analytics processes are combined.
Our goals were: 

\begin{itemize}[topsep=4pt, itemsep=4pt]
    \item Investigate if and how people use visual analytics operations such as encode or filter to help them integrate new attributes in-situ.
    \item Investigate if and how users' strategies and analytical behaviors during visual data analysis differ between ex-situ and in-situ integration approaches.
\end{itemize}

We built two interfaces with \add{different}\strike{ opposite} approaches to data preparation and analysis: the {\niitfc} interface, with manual file-based ex-situ integration as a separate step from visual analytics operations; and the {\iitfc} interface, with automatic UI-based in-situ integration merged with visual analytics operations.
Both simplify integration to column concatenation and selection to reduce confounds in observing users' analysis strategies and interactions.
This improved our ability to directly compare interactions and behaviors between these two different interfaces.
To further distinguish in-situ and ex-situ integration strategies in the {\iitfc} interface, we define \textit{primary} and \textit{secondary} integration processes and attribute interactions in Sect.~\ref{sec:combined_interface}.
We utilized a two (interfaces) by two (data sets) within-subjects experiment design, exposing participants to both interfaces and data sets to foster reflection on how their analysis process differed between conditions.
16 participants performed specific and free-form style visual data analysis tasks with each interface and data set in a random order.
We logged all mouse events and captured both the screen and audio of participants as they followed a think-aloud protocol \cite{Ericsson:1984:ProtocolAnalysis} and described their experience in a post-study semi-structured interview.
In this section, we describe the data sets and tasks curated, the experimental systems developed, and the procedure employed in this study.

\begin{figure}[!t]
  \centering
  \setlength{\abovecaptionskip}{0pt}
  \includegraphics[width=\linewidth]{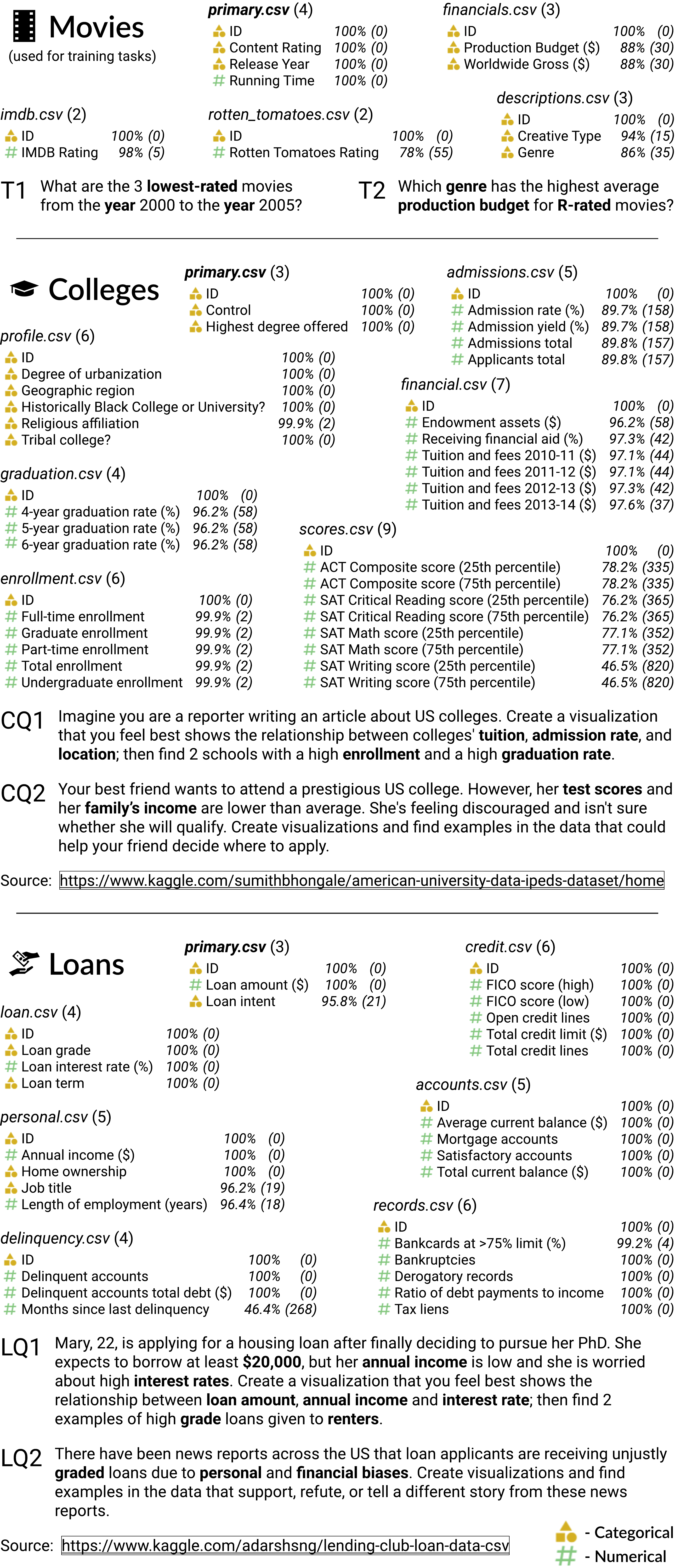}
  \caption{%
  Data sets and tasks used in the study (Sect.~\ref{sec:data_sets_and_tasks}).
  All data sets list each attribute's data type, percent complete (in italics), and total number of missing records (in parentheses), grouped under the files they are in.
  }%
  \label{fig:data_sets_and_tasks}
  \vspace{-1em}
\end{figure}

\subsection{Data Sets and Tasks}
\label{sec:data_sets_and_tasks}
We modified two publicly available tabular data sets on Kaggle relating to colleges and loans and generated a training movies data set in the style of the other two data sets, shown in Fig.~\ref{fig:data_sets_and_tasks}.
The modifications to the colleges and loans data sets include reducing the number of numerical attributes in the colleges data set and the number of records in the loans data set to: (1) make it easier to find categorical attributes; and (2) reduce the total number of data points to inspect visually, respectively.
\add{While these changes limited ecological validity, they also} reduced potential confounds in participant feedback when scaling integration to large data sets.

The final colleges data set contained all 1534 records from the original and comprised one primary key (\textit{ID}), seven categorical attributes, and 26 numerical attributes out of the 145 original attributes.
Each record of the colleges data set represents one U.S. college.
The final loans data set contained a subset of 500 records from the original and comprised one primary key (\textit{ID}), five categorical attributes, and 21 numerical attributes.
Each record of the loans data set represents one loan application.
The final movies data set contained 250 records and comprised one primary key (\textit{ID}), four categorical attributes, and five numerical attributes.
Each record of the movies data set represents one movie.
For all data sets, the attributes were distributed between the \textit{primary.csv} file loaded into the interface to start and all other CSV files, with a one-to-one mapping of records between every file using the attribute \textit{ID} as a relational key.
The records in each file except \textit{primary.csv} were randomly sorted, any values missing in the original data sets for a given record or attribute were kept, and any potentially identifiable information, such as the names of colleges or loan applicants, was removed.

We created two tasks for the colleges and loans data sets (\textit{CQ1}, \textit{CQ2}; and \textit{LQ1}, \textit{LQ2}), labeled by data set (\textit{C} and \textit{L})  and task type (\textit{Q1} and \textit{Q2}), as well as two specific training tasks (\textit{T1} and \textit{T2}) for the training movies data set.
The task descriptions for each data set can be read in Fig.~\ref{fig:data_sets_and_tasks}.
Participants were instructed to discover insights from the data by locating attributes of interest and browsing for patterns visually, then summarizing the uncovered relationships between visualized attributes \cite{Brehmer:2013:Multi-levelTypology}.
The first task of the colleges and loans data sets, henceforth called the \textit{specific task} and labeled \textit{CQ1} and \textit{LQ1}, required a specific visualization and data points to answer successfully.
The second task of the colleges and loans data sets, henceforth called the \textit{free-form task} and labeled \textit{CQ2} and \textit{LQ2}, featured open-ended analysis and assignments to interpret the data.
Every task description listed at least one relevant attribute that was not in \textit{primary.csv}, thus requiring participants to locate attributes in other files.
For example, in \textit{CQ1}, none of ``tuition'', ``admission rate'', ``location'', ``enrollment'', or ``graduation rate'' are located in \textit{primary.csv}.

\subsection{Experimental Systems}
\label{sec:experimental_systems}
We developed both {\niitfc} and {\iitfc} interfaces inspired by PoleStar \cite{Wongsuphasawat:2015:Voyager} (Fig.~\ref{fig:experimental_systems}).
With these interfaces, participants encoded data via drop-down menus and manipulated it via select, arrange, change, filter, and aggregate operations \cite{Yi:2007:InteractionTaxonomy, Brehmer:2013:Multi-levelTypology}.
The primary difference between these two interfaces is how they structure the data integration process.
The {\iitfc} interface has data integration functionality (column concatenation) embedded directly into drop-downs (column selection), traditional user interface (UI) controls used for visual data analysis found in tools such as Tableau.
In comparison, the {\niitfc} interface has the same visualization UI for column selection, but data integration happens outside of the interface, where users manually combine attributes into \textit{primary.csv} using Excel via column concatenation.

\subsubsection{Separated Interface}
\label{sec:separated_interface}
This baseline interface shown in Fig.~\ref{fig:experimental_systems} features six panels for performing visual data analysis: an Attributes panel for displaying the attributes and their data types from \textit{primary.csv}; an Encode panel for specifying the data-visual mapping; a Filter panel for applying categorical and numerical filters to the data; a Visualization panel that allows users to hover and click on the data marks; and an Elaborate panel that shows a table of the records corresponding with hovered- and clicked-on data marks.
A Navigation bar at the top of the screen shows participants the current task description.
To integrate data, users navigated the operating system's file browser to open CSV files in Excel that collectively contained the data sets used for the study (see Sect.~\ref{sec:data_sets_and_tasks} for a description of the data and Fig.~\ref{fig:data_sets_and_tasks} for a representation of the file structure).
Participants were allowed to concatenate columns between files using any operations available in Excel.
The results of their operations, including the integrated attributes, were loaded into the interface through a single file, \textit{primary.csv}, via the Refresh button in the Attributes panel.
We chose Excel as a baseline for manual data integration for two reasons: (1) it mirrors real decision-making scenarios described by Dimara and Stasko \cite{Dimara:2021:DecisionMakingTasks}; and (2) its familiarity in our target participant group.

\subsubsection{Combined Interface}
\label{sec:combined_interface}
\begin{figure}[!t]
  \centering
  \setlength{\abovecaptionskip}{0pt}
  \includegraphics[width=\columnwidth]{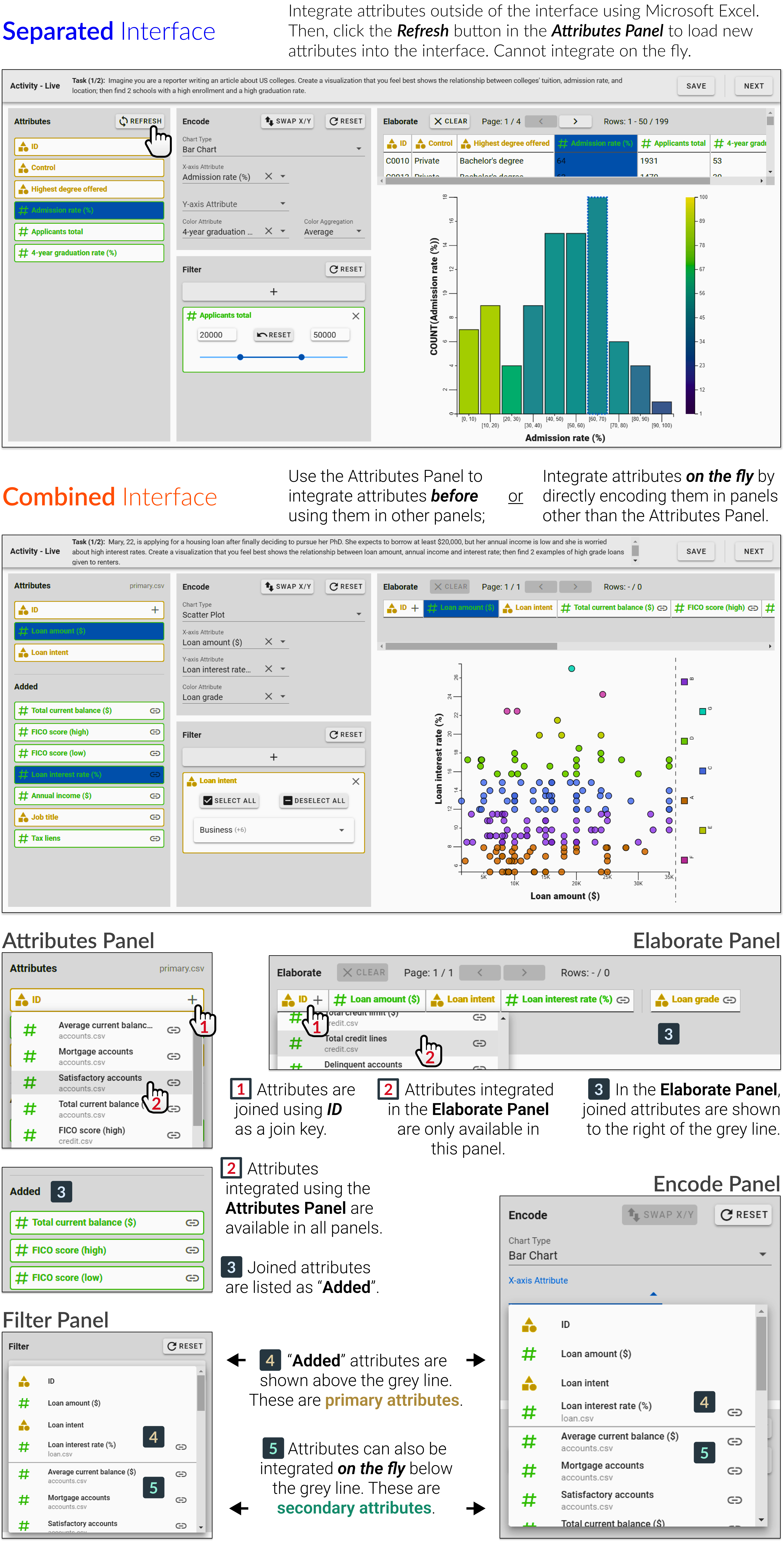}
  \caption{%
  Our two interfaces: (1) {\niitfc} (top; showing task \textit{CQ1}); and (2) {\iitfc} (bottom; showing task \textit{LQ1}) (Sect.~\ref{sec:experimental_systems}).
  }%
  \label{fig:experimental_systems}
  \vspace{-1em}
\end{figure}

In contrast to the {\niitfc} interface, we designed a second interface shown in Fig.~\ref{fig:experimental_systems} that incorporates integration operations directly into the Attribute, Encode, Filter and Elaborate panels without revisiting separate files or tools, e.g., Excel.
To help participants focus on how in-situ integration affected their analysis process, we sought to reduce potential confounds, such as join errors, that could arise in feedback.
We achieved this by merging the separate files for each data set used with the {\niitfc} interface into a single file and schema when using the {\iitfc} interface.
Then, when participants used a panel to add an attribute to their analysis, a lookup was performed.
By exposing the operation as a join in the interface and to the participant, we believe this study design decision helped keep feedback focused on how the combination of integration and visualization operations affected analysis.

\medskip
\noindent\textbf{Primary and secondary attributes. }
While we know that ex-situ integration operations can affect users' analysis processes during visual data analysis \cite{Pirolli:2005:SensemakingProcess}, it is unclear if analogous in-situ integration operations uniquely affect users' workflows and analytical behaviors.
To distinguish ex-situ and in-situ integration, we separated the \textit{primary} integration process of adding attributes into the Attributes panel, available in both interfaces, from the \textit{secondary} method of integrating attributes on the fly in the Encode and Filter panels, available only in the {\iitfc} interface.
Thus, we define \textit{primary} attributes as those attributes that are present in the Attributes panel when they are used in any panel, and \textit{secondary} attributes are those attributes that are not present in the Attributes panel when they are used in any panel.
Fig.~\ref{fig:experimental_systems} describes the location of \textit{primary} and \textit{secondary} attributes relative to the {\iitfc} interface panels.
For example, a user could visualize an attribute from outside \textit{primary.csv} on the fly in the Encode panel (i.e. \textit{secondary} attribute interaction), then intentionally add it to \textit{primary.csv} and their Attributes panel and encode it once again (i.e. \textit{primary} attribute interaction).
They could also mostly integrate and visualize attributes on the fly (i.e. \textit{secondary} attribute interactions), a unique integration strategy, or mostly integrate attributes before visualizing them (i.e. \textit{primary} attribute interactions), similar to the {\niitfc} interface.
We describe the results from investigating these interactions in Sect.~\ref{sec:interactions}.

\medskip
\noindent\textbf{Attributes panel. }
In the Attributes panel (Fig.~\ref{fig:experimental_systems}), participants can integrate attributes into  \textit{primary.csv} via a drop-down menu accessed by clicking on the join key (\textit{ID}).
Each item in the menu is an attribute labeled with data type, source file, and an indicator that the attribute is not originally from \textit{primary.csv}.
Once clicked, attributes are joined into \textit{primary.csv} and displayed under the ``Added'' header.

\medskip
\noindent\textbf{Encode and Filter panels. }
In the drop-down menus of the Encode and Filter panels (Fig.~\ref{fig:experimental_systems}), ``Added'' attributes from the Attributes panel can be used directly (i.e. \textit{primary} attribute interactions).
Attributes not in \textit{primary.csv} can also be used (i.e. \textit{secondary} attribute interactions) and thus ``seamlessly'' integrated on the fly through visual analytics operations.
``Added'' attributes are shown above the horizontal dividing line while all other attributes available to integrate on the fly are shown below it.
Integrating an attribute on the fly does not add it to the Attributes panel; however, once an attribute is added to the Attributes panel, it moves above the dividing line and can no longer be integrated on the fly.

\medskip
\noindent\textbf{Elaborate panel. }
In the Elaborate panel (Fig.~\ref{fig:experimental_systems}), participants access a drop-down menu by clicking on the join key (\textit{ID}) to add attributes from separate files directly to that panel without adding them to the Attributes panel, similar to the Encode and Filter panels.
Attributes added on the fly are shown to the right of the vertical dividing line.
If an attribute added on the fly is then added to the Attributes panel, it moves to the left of the vertical dividing line.

\subsection{Study Procedure}
\label{sec:study_procedure}
We recruited 16 participants (P$1-16$) via recruitment emails to university mailing lists.
Seven self-identified as male, and nine self-identified as female.
All of the participants held or were pursuing undergraduate or higher degrees in fields spanning Computer Science (8), Analytics (4), Human-Computer Interaction (2), Human-Centered Computing (1), and Industrial Design (1).
Participants self-reported prior engagement with a wide range of visual data analysis tools including Tableau (15), Python/Matplotlib (11), R/ggplot2 (6), Microsoft Power BI (4), D3.js (2), SAS (2), and AWS Quicksight (1).
All participants also had prior experience with using Excel for visual data analysis: 11 participants self-reported moderate experience; the rest self-reported as having either a lot or a little experience.

After obtaining consent, participants self-reported their demographics in a pre-study survey.
Then one of each of the two possible interfaces and data sets was randomly chosen to control for ordering effects.
The researcher conducting the session explained aloud the various features of the interface.
Participants were then asked to complete a training task (\textit{T1} or \textit{T2}) with the movies data set that would not be evaluated.
After, they were asked to complete a specific (\textit{Q1}) task, then a free-form (\textit{Q2}) task, that would both be evaluated.
Participants then had the other interface explained aloud to them and were similarly asked to complete a training task (\textit{T1} or \textit{T2}) with the movies data set, a specific (\textit{Q1}) task, and finally a free-form (\textit{Q2}) task using the other data set and the other interface.
Participants were numbered in this paper according to this counterbalancing effort: participants P$1-8$ used the {\iitfc} interface with the loans data set and the {\niitfc} interface with the colleges data set; and participants P$9-16$ used the {\iitfc} interface with the colleges data set and the {\niitfc} interface with the loans data set.
Finally, the participant and researcher engaged in a semi-structured debrief interview discussing the participant's experience during the session.

We asked participants to use a think-aloud protocol \cite{Ericsson:1984:ProtocolAnalysis} as they worked and to summarize how they arrived at their results after completing the tasks.
We recorded the screen and audio (participant and researcher) as well as interaction logs of all mouse events.
Each session took place in-person and lasted between two and three and a half hours, with a mean time of two and a half hours.
Each participant was compensated \$30 USD via an Amazon gift card.

\section{Study Results}
\label{sec:results}

\begin{figure}[!t]
  \centering
  \setlength{\abovecaptionskip}{0pt}
  \includegraphics[width=\columnwidth]{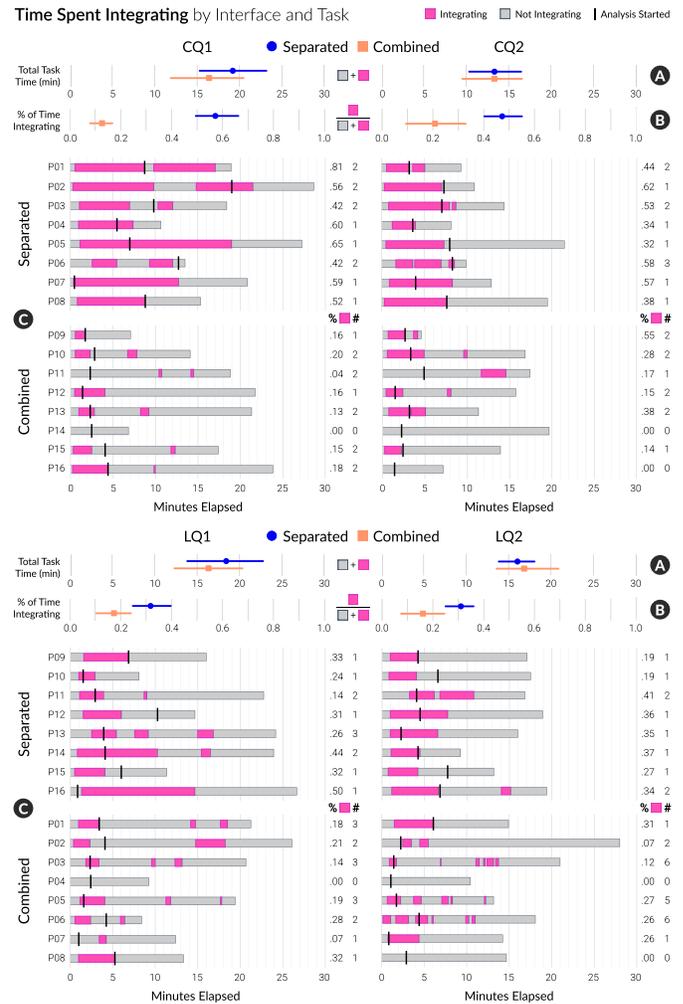}
  \caption{%
  Bootstrapped 95\% CIs around the mean estimations of total task completion time \textbf{(A)} and percent time spent integrating between interfaces \textbf{(B)}, as well as the time spent integrating organized by interface and task \textbf{(C)} (Sect.~\ref{sec:time_spent}).
  Each estimate represents eight participants with 1000 resamples.
  The darker pink bars represent intervals of data integration and the vertical black lines are a proxy for showing when analysis started.
  The percent (\%) of time spent integrating and number (\#) of intervals of integration are shown to the right of each bar.
  }%
  \label{fig:time_spent_integrating}
  \vspace{-1em}
\end{figure}

\begin{figure*}[!t]
  \centering
  \setlength{\abovecaptionskip}{0pt}
  \includegraphics[width=\linewidth]{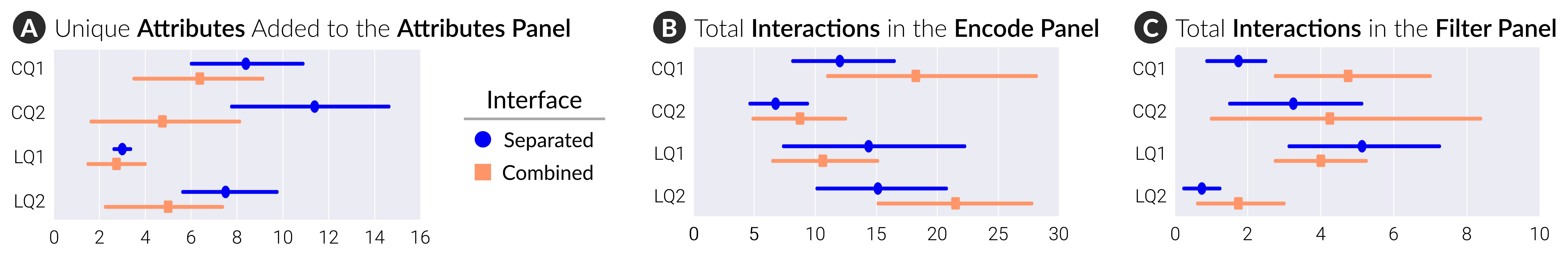}
  \caption{%
  Bootstrapped 95\% CIs around the mean estimations of \textbf{(A)} unique attributes added to the Attributes panel as well as total attribute interactions between interfaces in the \textbf{(B)} Encode and \textbf{(C)} Filter panels (Sect.~\ref{sec:interactions}).
  Each estimate represents eight participants with 1000 resamples.
  }%
  \label{fig:total_int_attrs}
  \vspace{-1em}
\end{figure*}

We used event logs and video recordings to uncover quantitative and qualitative patterns in participant's interactions and behaviors.
Following Dragicevic \cite{Dragicevic:2016:HCIStats}, we interpreted effect sizes such as sample means using bootstrapped 95\% confidence intervals (CIs) with 1000 resamples to represent uncertainty (Fig.~\ref{fig:time_spent_integrating}, Fig.~\ref{fig:total_int_attrs}).
For a given confidence level and sample size, CI width increases with increasing variability; results are generally significant if CIs do not overlap.
We also provide absolute counts as bar charts (Fig.~\ref{fig:time_spent_integrating}, Fig.~\ref{fig:primary_secondary_interactions}).
We further evaluated both the video recordings of each session and the audio recordings of participants' think-aloud protocol, debrief interview, and the researcher's notes, and conducted inductive thematic analysis \cite{Boyatzis:1998:ThematicAnalysis}, identifying emergent themes that were discussed amongst all authors.
We acknowledge that a think-aloud protocol carries the potential to affect time spent on tasks and interactions during studies and consider this in our subjective interpretations.
In this section, we present preliminary observations and open questions from our mixed-methods analysis around time spent (Sect.~\ref{sec:time_spent}), interactions (Sect.~\ref{sec:interactions}), integration strategies (Sect.~\ref{sec:integration_strategies}), and analytical behaviors (Sect.~\ref{sec:analytic_behaviors}).

\subsection{Time Spent}
\label{sec:time_spent}
When presented with both in-situ and ex-situ integration capabilities, will participants spend time differently between interfaces and tasks?

We needed to temporally separate when participants were integrating and visualizing data, especially in-situ where these operations are combined.
From Pirolli and Card \cite{Pirolli:2005:SensemakingProcess} we know that users locate sources of data, integrate these data, and then analyze them, in a foraging and sensemaking loop.
We designed both of our interfaces with an Attributes panel for users to keep track of what they have integrated in the loop.
Thus, as a first step, we can use the set of operations that result in an attribute being added to the Attributes panel as a proxy for intervals of data integration in both interfaces.
In the {\niitfc} interface, it is the time between when Excel is first made the top-level, primary window on the screen, at least one attribute is added to the Attributes panel, and the next time that the participant creates or modifies a visualization.
In the {\iitfc} interface, it is the time spent between when the Attributes panel drop-down menu is first opened, at least one attribute is clicked on, and the next time that the participant creates or modifies a visualization.
We also defined a proxy for when the participant started their analysis as the moment they first assigned an attribute to an encoding channel in the Encode panel.
We reviewed our video recordings and applied these definitions when manually recording the time intervals of integration.

\medskip
\noindent\textbf{Observations. }
Comparing CI width and overlap of task completion time in Fig.~\ref{fig:time_spent_integrating}A, we found no significant difference between interfaces or tasks.
The percentage of time spent integrating in Fig.~\ref{fig:time_spent_integrating}B was significantly different between interfaces; between 25\% to 65\% with the {\niitfc}, while only 10\% to 30\% with the {\iitfc}.
P$7$ described this difference during their session: \textit{``In the first [{\niitfc}] interface, I had to go through 8 different files to select the attributes, sort them, and add them to primary.csv. That was tedious and difficult compared to the second [{\iitfc}] interface, where those operations were much smoother.''}

Breaking down when participants integrated throughout the task in Fig.~\ref{fig:time_spent_integrating}C, we see different user behaviors emerging.
When using {\niitfc} interface, participants mostly integrated in one or two intervals and usually before starting analysis, whereas with the {\iitfc} interface, integration usually happened in two or more intervals any time throughout the analysis.
For P$9$, using the {\niitfc} interface mirrored processes they followed with other visual analytics tools: \textit{``When I try to work on visualization, I think of it as a two-step process: I find the attributes first, then make the visualizations. Otherwise it's a lot to keep track of and think about... I'm just in the habit of making my list before visualizing... I think of the tasks as separate... I think my experience in Tableau makes me expect to have to connect data in sheets first.''}
Unique strategies also emerged; e.g., P$4,8,14,16$ exclusively integrated attributes on the fly (i.e. no time spent integrating and no pink bars in Fig.~\ref{fig:time_spent_integrating}C), while P$3,5,6$ used the Attributes Panel to integrate one attribute at a time up to 6 times (i.e. many small pink bars in Fig.~\ref{fig:time_spent_integrating}C).
P$8$ switched to this strategy during the free-form task (\textit{Q2}) saying: \textit{``It was very quick and efficient to test and move attributes.''}
Differences between task types (\textit{Q1} and \textit{Q2}) were inconclusive.

\add{When should data integration operations be combined with visual analytics operations?}
Overall, we were surprised to see that both specific and free-form tasks took roughly the same time to complete with either interface, even though there are fewer integration operations to perform in the {\iitfc} interface.
We also saw participants spend more time integrating data before analyzing it using the {\niitfc} interface, whereas integration happened less often and more frequently throughout analysis with the {\iitfc} interface.
This could suggest that in-situ integration helps users stay focused on analysis tasks longer than ex-situ integration.
At the same time, some participants chose when and how long to integrate regardless of total task time based on previous experience with tools like Tableau, where integration usually happens separately from analysis.
Users may require a balance between manual and automated integration.

\medskip
\noindent\textbf{Open questions. }
In addition to the findings above, our study revealed open questions, listed below, that encourage future research in this area:

\begin{enumerate}[topsep=4pt, itemsep=4pt]
  \item Are users spending more time thinking about their analysis when using in-situ versus ex-situ integration?
  \item Are users able to quickly transition between integration and analysis with in-situ integration, or does new data being introduced interrupt the flow?
  \item What factors in a user's prior experience affect time spent? How do they correlate with task requirements and interface design?
  \item Will time spent be affected if users can choose between in-situ and ex-situ integration on the fly?
\end{enumerate}

\subsection{Interactions}
\label{sec:interactions}
Will in-situ or ex-situ integration reveal differences in how users perform visual analytics operations?

\medskip
\noindent\textbf{Observations. }
Comparing CI width and overlap of estimations in Fig.~\ref{fig:total_int_attrs}, we found participants integrated slightly more unique attributes into their analysis using the {\niitfc} interface overall (Fig.~\ref{fig:total_int_attrs}A), yet interacted with slightly more attributes in both the Encode and Filter panels of the {\iitfc} interface (Fig.~\ref{fig:total_int_attrs}B and Fig.~\ref{fig:total_int_attrs}C), though not enough to be significant.
P$15$ demonstrated both of these patterns by (1) mostly using the Attributes panel to integrate and organize lots of unique attributes (\textit{``Since there were a lot of attributes, having them at the top of the drop-downs was useful''}) while (2) also exploring new attributes on the fly in the Encode and Filter panels because it was easy (\textit{``In the [{\iitfc}] interface, I didn't have to add the attribute to the Attributes panel, I could just add the attribute to the panel I wanted it in''}).

\begin{figure}[!t]
   \centering
   \setlength{\abovecaptionskip}{0pt}
   \includegraphics[width=\linewidth]{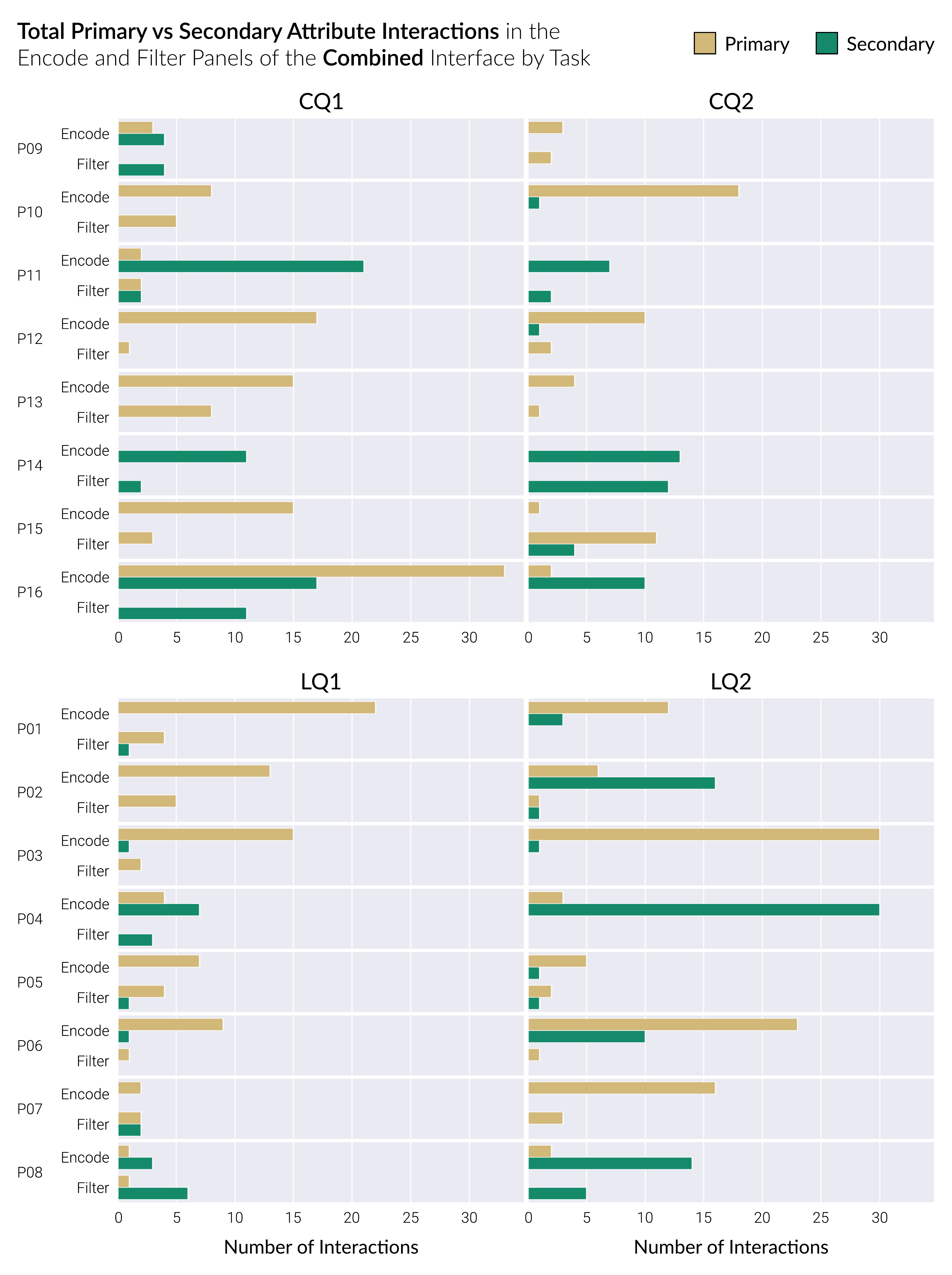}
   \caption{%
   Total counts of \textit{primary} and \textit{secondary} attribute interactions in the Encode and Filter panels of the {\iitfc} interface (Sect.~\ref{sec:interactions}).
   See Sect.~\ref{sec:combined_interface} and Fig.~\ref{fig:experimental_systems} for definitions.
   }%
   \label{fig:primary_secondary_interactions}
   \vspace{-1em}
\end{figure}

Based on whether attributes were integrated ex-situ (\textit{primary}) or in-situ (\textit{secondary}) in the {\iitfc} interface, we observed three distinct patterns of interaction in Fig.~\ref{fig:primary_secondary_interactions} and compared them with the intervals of data integration in the {\iitfc} interface (Fig.~\ref{fig:time_spent_integrating}C).
$9/16$ participants (P$1,3,5,6,7,10,12,13,15$) interacted mostly with \textit{primary} attributes, similar to how the {\niitfc} interface is used; $6/9$ of these participants (P$1,7,10,12,13,15$) also spent a majority of their time integrating with the Attributes panel before starting their analysis.
$4/16$ participants (P$4,8,11,14$) interacted mostly with \textit{secondary} attributes; $3/4$ of these participants (P$4,11,14$) also rarely used the Attributes panel, in contrast to how the {\niitfc} interface is used.
$3/16$ participants (P$2,9,16$) interacted with a mixture of \textit{primary} and \textit{secondary} attributes with no clear preference.
P$9$ based their interactions on the task: \textit{``If it's a specific task and we need attributes `x', `y', and `z', then we can directly use the Encode panel. But if I have a free-form task, then I want to shortlist my attributes first [in the Attributes panel] and explore those.''}
P$2$ initially used the Attributes panel, then switched to integrating with the Encode and Filter panels, saying \textit{``I don't think adding attributes to the Attributes panel made a difference... that is not where I use my attributes.''}
Attribute panel usage was inconsistent for these three participants and the four remaining participants (P$3,5,6,8$) with mostly \textit{primary} or \textit{secondary} attribute interactions.
For example, P$8$ spent time using the Attributes panel before starting analysis in \textit{LQ1}, then did not use the Attributes panel at all in \textit{LQ2}.
Yet we found that they mostly interacted with \textit{secondary} attributes in the Encode and Filter panels across both tasks.

\pagebreak

\add{Where should data integration operations be combined with visual analytics operations?}
Overall, while attributes \add{took less operations}\strike{ were easier} to integrate in the {\iitfc} interface, some participants integrated slightly more unique attributes with the {\niitfc} interface.
We also did not see large differences in total interactions with attributes in either the Encode or Filter panels between interfaces.
In-situ integration \add{taking less operations}\strike{ being easier} may not strongly affect how many attributes users interact with compared with ex-situ integration, suggesting a more balanced approach between in-situ and ex-situ.
Yet considering that the Attributes panel is a common way to incorporate new attributes during analysis and is our proxy for intervals of data integration in both interfaces, we were surprised that Attributes panel usage in the {\iitfc} interface was split.
Similarly, while \textit{primary} attribute interactions are more common in tools like Tableau, \textit{primary} and \textit{secondary} attribute interactions in the {\iitfc} interface were also split and users did not converge on a single strategy for using integrated attributes during analysis.
Participants further demonstrated preferences, e.g., those with a preference towards \textit{primary} attribute interactions used the Attributes panel more often, and analogously less often for those with more \textit{secondary} attribute interactions.
Participant preferences for integration strategies (Sect.~\ref{sec:integration_strategies}) suggest that in-situ integration helps users during visual data analysis in unique ways, despite little difference in interactions.
We describe participants' strategies for using in-situ integration in the next section.

\medskip
\noindent\textbf{Open questions. }
In addition to the findings above, our study revealed open questions listed below:

\begin{enumerate}[topsep=4pt, itemsep=4pt]
  \item If users can choose between in-situ and ex-situ integration on the fly, will constraints based on time, task requirements, or interface design affect users' interactions with a large coverage of attributes?
  \item Similar to attribute coverage, if \add{less operations}\strike{ easier integration} does not lead to more interactions overall, will other constraints influence integration method preference?
  \item Are users adapting their analysis process to new ways of integrating on the fly? How does prior experience affect whether users choose in-situ or ex-situ integration?
\end{enumerate}

\subsection{Integration Strategies}
\label{sec:integration_strategies}
\add{How should data integration operations be supported in tandem with visual analytics operations?}
Based on participants' time spent and interactions, as well as video recordings and qualitative feedback, we identified four distinct integration strategies across both interfaces.
Most participants used a single strategy for each interface across both tasks, while a few switched strategies between tasks.

\medskip
\noindent\textbf{S1 -- ``Integrate attributes first'' -- both interfaces. }
Several participants integrated attributes before starting analysis and rarely integrated more afterwards.
\textbf{S1} is characterized by more time spent integrating attributes into the Attributes panel than other strategies and, in the {\iitfc} interface, more \textit{primary} attribute than \textit{secondary} attribute interactions.
The time spent adding attributes to the Attributes panel was usually apportioned to one integration session in the beginning of each task collecting a large subset of attributes, with participants rarely integrating after their initial collection.
With the {\niitfc} interface, this strategy was used almost exclusively since attributes could only be added to the Attributes panel via Excel.
When using the {\iitfc} interface, $7/16$ participants (P$1,7,8,9,12,15,16$) utilized this strategy at least once and $4/16$ used it exclusively (P$7,9,12,15$).
P$1$ used this strategy to make a plan before analyzing the data: \textit{``Based on the question, I didn't need to worry about all attributes. I wanted to pick attributes that would help me answer the question... I wanted to have in my head, 'what makes sense to visualize?', before starting analysis.''}

\medskip
\noindent\textbf{S2 -- ``Don't think about integrating attributes'' -- }{\iitfc}\textbf{ interface only. }
Some participants integrated on the fly and rarely shortlisted attributes before using them.
\textbf{S2} is characterized by little to no time spent adding attributes to the Attributes panel at any point during analysis and, in the {\iitfc} interface, mostly \textit{secondary} attribute interactions.
Instead, participants used attributes directly in the panels they were interested in; e.g., the Encode or Filter panel.
\textbf{S2} was not possible when using the {\niitfc} interface.
When using the {\iitfc} interface, $6/16$ participants (P$2,4,8,11,14,16$) utilized this strategy at least once and $3/16$ used it exclusively (P$4,11,14$).
P$6$ reflected on how they would use this strategy after the study: \textit{``Adding attributes in the Attributes panel was confusing. At first I used it to see what attributes were there, but you could do the same operation in the other panels... The more you know the attributes, the less you would need the Attributes panel.''}

\medskip
\noindent\textbf{S3 -- ``Integrate attributes as needed'' -- both interfaces. }
A few participants used their analysis process to inform them when to integrate, instead of all up front or on the fly.
\textbf{S3} is characterized similarly to \textbf{S2}, including more time spent adding attributes to the Attributes panel and, in the {\iitfc} interface, mostly \textit{primary} attribute interactions.
However, that time was apportioned more evenly throughout the task in an ``analyze as you go'' fashion, with participants returning two or three times to integrate smaller subsets of attributes as needed.
While this strategy was possible with the {\niitfc} interface, no participants engaged with it when using the {\niitfc} interface.
When using the {\iitfc} interface, $6/16$ participants (P$1,2,5,6,10,13$) utilized this strategy at least once and $2/16$ used it exclusively (P$10,13$).
P$13$ describes this strategy as an extension of their analysis process: \textit{``I do that normally. I wanted to make changes then see what happens... what if I come across something interesting as the visualization changes?''}
For them, integrating helps them generate insights: \textit{``I think the free-form task was more about getting insights. I was seeing if adding or deleting anything was changing the visualization.''}

\medskip
\noindent\textbf{S4 -- ``Integrate attributes one-at-a-time'' -- both interfaces. }
A handful of participants focused on a small number of attributes in detail, adding them to the interface one at a time.
\textbf{S4} is characterized by less time spent adding attributes to the Attributes panel than the other strategies and, in the {\iitfc} interface, mostly \textit{primary} attribute usage.
To achieve this, participants briefly and sporadically added a few attributes to the Attributes panel as many as six times throughout a task.
Only P$13$ used this strategy with the {\niitfc} interface, while working on LQ1, returning to Excel multiple times for integrating one to two attributes at a time.
When using the {\iitfc} interface, $3/16$ participants (P$3,5,6$) utilized this strategy at least once and one used it exclusively (P$3$).
P$3$ used this strategy to organize their Attributes panel: \textit{``At first, I tried putting in the specific attributes into the Encode panel, but that wasn't such a good idea. I want to have all of the attributes organized together in the Attributes panel instead.''}
P$5$ used it to organize the Elaborate panel: \textit{``I wanted to see the order of attributes I added in the Attributes panel in the Elaborate table as well.''}

\medskip
\noindent\textbf{Switching strategies. }
Some participants changed their strategies between the specific task (\textit{Q1}) and free-form task (\textit{Q2}).
It is not clear what effects the task type had on why participants switched tactics.
Four participants (P$1,2,5,6$) started with \textbf{S3}: P$1$ switched to \textbf{S1}; P$2$ switched to \textbf{S2}; and P$5,6$ switched to \textbf{S4}.
Two participants (P$8,16$) started with \textbf{S1}, then both switched to \textbf{S2}.

\medskip
\noindent\textbf{Ex-situ integration operations.}
Two ways of integrating attributes ex-situ emerged: (1) copy-and-paste; and (2) the Excel function \textit{V-LOOKUP}.
$13/16$ participants copied and pasted columns from one file to the next.
P$15$ explains that \textit{``it was similar to how I use other data analysis tools. I would copy and paste data from one spreadsheet to another.''}
Participants varied in whether and how they would validate the success of their joins.
For example, P$11$ told us that \textit{``when I was copy-pasting, I assumed that the rows from the primary table were all there in the secondary tables.''}
$3/16$ participants (P$2,12,16$) used \textit{V-LOOKUP} to populate a new column with values based on a look-up with the primary key (\textit{ID}) column, ensuring a correct join.
P$1$ mentioned using \textit{V-LOOKUP} as preferable to copy-and-paste but did not use it, citing a lack of time and familiarity: \textit{``Sorry it's a bit slow [copy-and-paste], I'm not very good at Excel... another good way [to integrate] would be to use a V-LOOKUP to match the IDs... But looking at the IDs, they look like they match up.''}

\subsection{Analytical Behaviors}
\label{sec:analytic_behaviors}
\add{How will incorporating data integration into an on-going visual analytics process affect user behaviors?}
\add{We describe observed analytical behaviors related to satsificing and exhibiting bias from a sensemaking perspective.}\strike{ Participants also demonstrated analytical behaviors related to satsificing and exhibiting bias. We relate these behaviors to the analysis process from a sensemaking perspective.}

\medskip
\noindent\textbf{Satisficing. }
We observed patterns of satisficing, a cognitive heuristic for choosing a satisfactory or ``good enough'' option from alternatives \cite{Heuer:1999:IntelligenceAnalysis}.
Pirolli and Card explain that ``time pressures and data overload work against the individual analyst's ability to rigorously follow effective methods for generating, managing, and evaluating hypotheses'' \cite{Pirolli:2005:SensemakingProcess}.
P$5$ managed the constraints of integration by prioritizing insight generation: \textit{``I got less time to decide on which attributes to use, and I spent more time on the data pre-processing. I would prefer the [{\iitfc}] interface more. In visual data analysis, it's more important to gain insights.''}
P$9$ told us that they considered not introducing attributes into their analysis as a consequence of their sensemaking: \textit{``It was a lot of operations to just add a single variable. If I was 50/50 about whether to include an attribute, then I may not include it in my analysis.''}

Conversely, from a data-frame theory perspective, Klein et al. suggest that sensemaking is the balance of fitting data to a frame and the frame affecting how data is interpreted \cite{Klein:2007:DataframeSensemaking}.
When asked why they satisficed, P$9$ attributed the difference between interface designs to their trust in the data: \textit{``In terms of accuracy and insights the [{\niitfc}] interface was better. But the simplicity of the [{\iitfc}] interface was better... I think it all comes down to how much you trust the data. If you trust it, the [{\iitfc}] is better. But if the data isn't clean, the [{\niitfc}] is better.''}
Having integration be a simpler and more seamless part of visual data analysis could improve both the generation and coverage of hypotheses afforded by access to new attributes.
At the same time, removing the seams may affect the balance of fitting data to frames and frames to data that proceeds under the constraints of time pressures and data overload.

\medskip
\noindent\textbf{Exhibiting bias. }
We identified potential examples of cognitive bias due to the separation of data preparation and analysis.
For example, participants visualized the same set of attributes in similar ways, often saying they were ``familiar'' with these attributes, and may have been exhibiting confirmation bias \cite{Wall:2018:HumanBiasInVA}.
We expected participants to consider using different combinations of attributes to change their perspective, particularly considering the ease of immediately viewing integrated attributes in the {\iitfc} interface.
However, many of these participants instead claimed that ``the data just wasn't showing them what they wanted to see'' and continued attempting to preserve their frames by rearranging the encodings of the same attributes.
Klein et al. suggest that initial anchors, in this case the data that these participants first integrated, can have profound effects on performance during sensemaking tasks \cite{Klein:2007:DataframeSensemaking}.

Similarly, we found a tendency for some participants to exhibit anchoring effects \cite{Cho:2017:AnchoringEffectInVIS, Wall:2018:HumanBiasInVA} by sticking to a smaller number of attributes for a longer amount of time before attempting to branch out and find more information, if at all.
Often these participants, when asked how their analysis was going, would tell us that they were ``trying to make it work''.
However, they rarely used the integration features of either interface to overcome these issues, instead focusing on trying different combinations of encodings and filters to solve the problem.
Both of these examples suggest that such biases could persist even when the interface design affords the opportunity to integrate data with a single click.

\section{Preliminary Guidelines}
\label{sec:guidelines}

Towards preliminary guidelines, many participants cited commonly raised concerns around affordances, direct manipulation of the data \cite{Hutchins:1985:DirectManipulationInterfaces}, and fluid interaction \cite{Elmqvist:2011:FluidInteraction}, such as the responsiveness of the system and lack of load times, the layouts of the drop-down menus and panels, the usefulness of dynamic querying \cite{Ahlberg:1992:DynamicQuerying}, and the user experience.
With these in mind and our findings, we synthesized three guidelines for designing future visual analytics interfaces that can support integrating attributes throughout an active analysis process: (1) show where and how data are being integrated (Sect.~\ref{sec:guideline_show_data}); (2) use in-situ integration for exploring the space of attributes (Sect.~\ref{sec:guideline_when_insitu}); and (3) balance manual and automated approaches (Sect.~\ref{sec:guideline_balance_approaches}).

\subsection{Show where the data comes from}
\label{sec:guideline_show_data}
The transparency of \textit{how} and \textit{what} data are integrated is essential for in-situ data integration within a visual analytics system.
For example, P$1,8,12$ all used strategy \textbf{S1} in the {\iitfc} interface and asked for more access to the raw data.
P$1$ specifically wanted access to verify the quality of the data: \textit{``I liked [{\iitfc}] for the ease of use, but I also like to see the actual data in the [{\niitfc}]. I would go back and use the tool to verify the quality of the data.''}
On the other hand, P$5$ felt that the {\iitfc} interface would process the raw data better than they could, based on their experience with similar visual analytics tools: \textit{``When I copy-and-pasted data in the [{\niitfc}] interface, I had to manage column names and there couldn't be manual errors, and I feel like the [{\iitfc}] interface would do a better job of overcoming those... I've used Tableau and Power BI. I was told that those interfaces mapped data correctly, so I assumed this one did as well.''}
While our study intentionally controlled for data quality, future interfaces should clarify the limitations of how and what data are integrated.
For example, Cashman et al. use a pop-up window in their CAVA system \cite{Cashman:2020:CAVA} to display how a join will be performed before it is ultimately integrated into the data set.
Analysis outcomes that follow from ``anonymous'' integration could be dangerous if not carefully evaluated.

However, we found strong evidence that showing the user all of what attributes can be integrated can negatively affect analysis in several ways.
For example, P$2$ expressed concerns of cognitive overload from having too many attributes to think about: \textit{``There were so many variables [in the drop-downs] that I missed some. If the attributes were laid out [in the Attributes panel] like in Tableau then maybe I would have seen it...''}
Too many attributes can also negatively affect task completion time.
With the {\iitfc} interface, P$10$ cited too many attributes in the drop-downs: \textit{``In the [{\iitfc}] interface, having too many attributes in the drop-downs in the Encode and Filter panels took more time than expected to look through.''}
With the {\niitfc} interface, P$5$ took time to traverse multiple files: \textit{``The attributes were spread across a lot of files. Traversing the files to find the right attribute took a while.''}
Further, making attributes more visible may contribute to satisficing.
P$11$ satisficed based on the task: \textit{``Sometimes I wasn't sure where to look... When doing the tasks, I would often look for just the attributes I felt like were relevant to the task. I ignored the rest because I had to go through the tables to find them.''}
P$6$ attributed this to interface design: \textit{``With the specific questions, I knew where to look in the tables with the [{\niitfc}] interface, compared with the [{\iitfc}] interface which was harder for this.''}
However, when the number of attributes felt manageable, P$9$ liked the drop-downs for finding relevant attributes: \textit{``I didn't know where the attributes were in the system, but that wasn't a problem. I had to skim through the attributes in the drop-down [of the {\iitfc} interface], but the number of attributes was not so great. If there was more, it would have been more difficult.''}
Thus, designers should carefully consider how the number of attributes in the data may influence users' time spent, interactions, and analytical behaviors by balancing how much is shown to the user at once with how easy it is for users to find relevant attributes to integrate during analysis.

\subsection{Use in-situ integration for exploration}
\label{sec:guideline_when_insitu}
The issues faced by participants using the {\niitfc} interface are common to data integration as an entry point to analysis.
There is some evidence that the overhead cost of integration outside the interface could prevent users from finding relevant attributes.
For example, P$2$ described trading off between analyzing and integrating, causing conflicts where the interface lacked support.
They attributed mistakes they made to managing multiple tables manually across several windows with the {\niitfc} interface: \textit{``I copied the values into the wrong file because so many windows were open. That wasted my time.''}
Instead, we saw more evidence that encode and filter operations were useful for exploring the attribute space to find and integrate new attributes on the fly in the {\iitfc} interface.
P$11$ found the grouping of attributes in a single drop-down helpful because it allowed them to see all of the available data: \textit{``I liked having all of the attributes that were relevant to the tasks in one place.''}
P$14$ felt the {\iitfc} interface helped them find more attributes: \textit{``It was more convenient to see all of the attributes in the [{\iitfc}] interface. For example, what if there was an attribute hiding in a table that I missed?''}
P$5$ used the names of the files as a proxy for determining the semantic relevance of an attribute to the task: \textit{``I didn't know what all attributes were [in the {\iitfc} interface], but I checked the names of the files for the attributes in order to choose which attributes to use.''}
Thus, in-situ data integration for quickly encoding new attributes in the visualization could help users maintain their focus on performing visual data analysis.
In the {\iitfc} interface, participants can immediately observe the results of integration in their visualizations and generate new ideas and strategies to explore \cite{Morton:2012:TableauDataBlending}.
This can also allow participants to evaluate the quality of the integrated information visually \cite{VanKleek:2013:CarpeData}.
An ``undo'' feature would further promote principles of direct manipulation that preserve the flow of the analysis process \cite{Elmqvist:2011:FluidInteraction}.
When combined with visual analytics operations, in-situ integration may help users maintain an active and continuous analysis process.

\subsection{Balance manual and automated approaches}
\label{sec:guideline_balance_approaches}
The differences in users' interactions and behaviors between ex-situ and in-situ integration provides some evidence for when manual approaches should be used over automated approaches.
P$2$ had difficulty remembering relevant attributes when the integration process was automated: \textit{``In Tableau and Power BI, you have to manually create tables when joining tables... but since I wasn't the one doing the joins [in the {\iitfc} interface], it was harder to remember the attributes that were available to me. I would have remembered them if I had to manually join the attributes.''}
P$16$ also preferred manual interactions for understanding their data: \textit{``I liked the [{\niitfc}] interface a lot even though it involved a lot of manual interactions with the CSVs... In the [{\niitfc}] interface, you are really visualizing your data, you understand how your data will be visualized. In the [{\iitfc}] interface, you have to understand how the different panels work together instead.''}
Yet there is evidence that automated approaches may improve the analysis process.
For example, despite tasks not being timed, the distribution of time spent with the {\niitfc} interface affected the performance of P$5$: \textit{``I think most of the time was spent doing data preparation and I felt rushed.''}
P$9$ also spent longer with the {\niitfc} interface because they had to separate analysis from integration: \textit{``It takes a long time to do manual integration. For example, when I open a CSV [file], I have thoughts about what it may contain, then I see the attributes. It's not the same operation to find the attribute and use the attribute, unlike in the [{\iitfc}] interface.''}
P$2$ elaborated on how maintaining context affected both their time spent and insights generated: \textit{``Adding attributes [to the {\niitfc} interface] was a pain. I had to open tables and copy and remove incorrect values. I had to look at the names of the files to guess where attributes would be... [my time] was mostly spent cleaning and arranging data. I didn't have a lot of time to focus on how I could improve the visualization.''}).
P$9$ explains how their experience differed between interfaces: \textit{``In the [{\iitfc}] interface, I can explore more attributes in a shorter amount of time. In the [{\niitfc}] interface, there's more overhead that takes time away from the task of visualizing data.''}
This suggests that designers should consider a minimal but fluid design \cite{Elmqvist:2011:FluidInteraction} for in-situ integration when time spent and interactions should be minimized, otherwise opt for manual approaches.
One potential in-situ solution could be the automation of integration steps that do not require as much human input.
Data blending techniques such as those provided by Tableau \cite{Morton:2012:TableauDataBlending} and Google Data Studio exemplify this idea by maintaining a human-in-the-loop control while abstracting away the more technical details of integration.
This may help users reduce the number of concurrent processes to manage while helping them maintain context.

\section{Limitations and Future Work}
\label{sec:limitations_future}

Beyond the preliminary state of our guidelines, we contribute open research questions and avenues for future work investigating more types of integration, task requirements, and users' prior experience towards ``concrete'' guidelines.

\medskip
\noindent\textbf{Types of integration. }
To elicit initial observations, we directly compared ex-situ and in-situ integration by intentionally simplifying and limiting integration to column concatenation and selection.
Because of this, we could not study technical challenges associated with more complex data integration, such as performing deduplication and entity resolution described in Sect.~\ref{sec:data_integration}, as well as issues of latency in how long it takes to resolve joins and data quality across heterogeneous sources.
Column concatenation itself is simple compared to ``real'' data integration that can involve extensive planning and even custom tool development when the semantics of data is complicated and/or involves semi-structured data such as text.
Additionally, we ensured a one-to-one mapping of recordings between every file and described all files and attributes to participants up front in Sect.~\ref{sec:data_sets_and_tasks}.
This choice did not allow us to investigate how the quality of data from potentially unknown sources affects users interactions and behaviors, e.g., introducing attributes from multiple websites or APIs, potentially on the fly as they become available, and throughout the analysis process as users request them.
Given these limitations, it is unclear whether users will consider the validity of a data source when data is not presented up front.
The way forward might require adopting (even developing) a more sophisticated approach to comparative evaluation not based on direct ``apples-to-apples'' comparison, e.g., focusing on single-table data wrangling with more complex integration tasks compared with equivalent field calculations wrapped in a UI-based approach.

\medskip
\noindent\textbf{Task requirements. }
We chose tasks following Fekete et al. \cite{Fekete:2008:ValueOfInfoVis}, who argue that visualizations are often best for exploratory tasks where the goal is to make discoveries or generate insights about data.
Our methodology did not allow us to investigate effects on task performance, including what markers indicate the end of a task and when or if participants decide to stop integrating.
Future work could model factors that determine how much data will be integrated and when based on task requirements and/or UI design.
It is unclear whether there is a threshold across which user interactions and behaviors change.
Additionally, while we found that our definition of time spent integrating closely matched qualitative feedback from participants, participants could have been using the time between integrating and visualizing data to consider what visualization to create that best utilizes the integrated attributes; this time may be significant.
We acknowledge that other measurements for marking the start and end of integration are also reasonable.
Finally, we limited the number of attributes in the data set, unlike real analysis scenarios where users may have \textit{a priori} knowledge of attributes they want and/or risk cognitive overload while browsing for relevant attributes out of many.
Our preliminary guidelines suggest not to overload users with every attribute that can be integrated; future work should isolate these effects and systematically describe them as factors in the data integration process during visual data analysis.

\medskip
\noindent\textbf{Users' experience. }
Our participants comprised a fairly homogeneous user group (all recruited from the same academic institution with varying levels of experience conducting data analysis, creating visualizations, and using data analysis software) to reduce confounds in comparing feedback.
Because of this, it is unclear whether incorporating data integration into visual analytics processes at all will depend on familiarity with software and analysis practices, particularly where data integration plays a major role, either as part of the analysis process or off-loaded to others.
For example, decision-makers with existing routines may or may not conduct analysis differently \cite{Dimara:2021:DecisionMakingTasks} when integration on the fly is possible.
Participants also varied in how much time they spent on tasks depending on their prior domain knowledge of the data sets, skill in performing visual data analysis, and comfort with learning and using the interfaces.
For example, P$16$ avoided attributes based on their domain experience: \textit{``In the loan question, I wasn't familiar with some of the terms, so I didn't really touch some of the files because I wasn't comfortable using them.''} 
Additionally, when participants felt that a task description was ambiguous or open to interpretation, they took longer to prepare data.
P$15$ reflected: \textit{``I feel like the difference in the quality of my analysis was less about the interface I was given and more about the task I was given.''}
Thus future studies should investigate how task requirements and existing domain knowledge impact integration during visual data analysis.
Our results indicate that it may be neither preferable nor realistic to start with a single file full of attributes.

\section{Conclusion}
\label{sec:conclusion}

This paper presents preliminary results and guidelines when combining data integration and visual analytics processes.
We developed two visual analytics interfaces: one that presents manual file-based ex-situ integration ({\niitfc}); and one that presents automatic UI-based in-situ integration ({\iitfc}).
We conducted a within-subjects user study with 16 participants and a mixed-methods analysis of participants' interactions and behaviors.

\add{Where and how should we support data integration operations in tandem with visual analytics operations?}
Participants spent time integrating before analysis, on the fly, and switching between strategies.
The time spent on tasks and interactions between interfaces was also not significantly different.
In-situ integration could enable analysts to explore attributes faster than analogous ex-situ strategies, leaving more time for analysis tasks.
\add{How will incorporating data integration into an on-going visual analytics process affect user behaviors?}
We observed participants using integration operations to generate and track hypotheses and insights as well as patterns of satisficing and bias in participants' analytical behaviors.
Supporting integration operations in visual analytics tools will also require transparency up front about what and how data are integrated as well as balancing both automated and manual approaches.

\ifCLASSOPTIONcompsoc
  \section*{Acknowledgments}
\else
  \section*{Acknowledgment}
\fi

This work was supported in part by the National Science Foundation grant IIS-1813281 and DRL-2247790.

\ifCLASSOPTIONcaptionsoff
  \newpage
\fi

\bibliographystyle{IEEEtran}
\bibliography{IEEEabrv,main}

\begin{IEEEbiography}[{\includegraphics[width=1in,height=1.25in,clip,keepaspectratio]{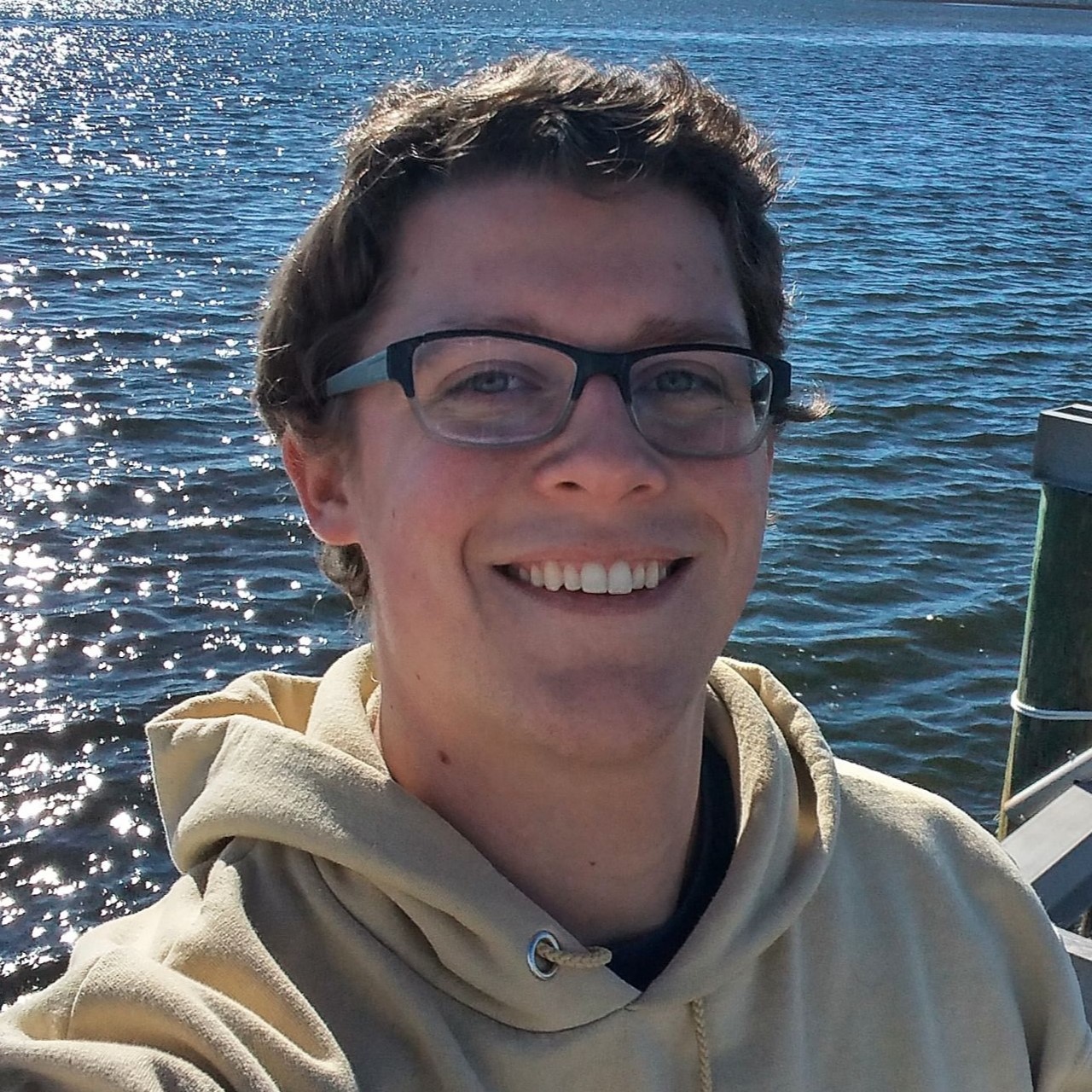}}]{Adam Coscia} is a PhD student at Georgia Tech’s School of Interactive Computing and a member of the Visual Analytics Lab. His research interests include Visual Analytics, Human-Computer Interaction, and Explainable Artificial Intelligence (AI) with Large Language Models and Knowledge Graphs. He received his B.S. in Physics. He won the President’s Fellowship for top incoming PhD students.\end{IEEEbiography}

\begin{IEEEbiography}[{\includegraphics[width=1in,height=1.25in,clip,keepaspectratio]{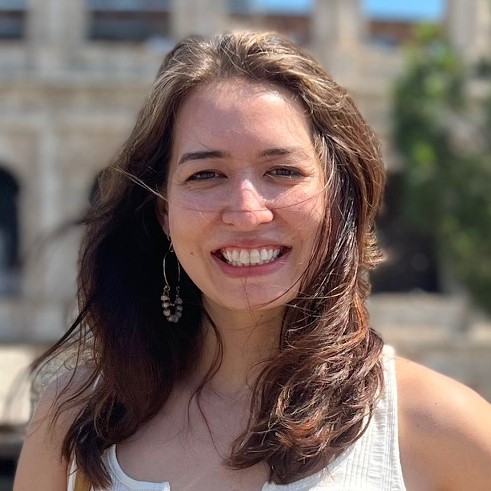}}]{Ashley Suh} received her PhD in computer science from Tufts University. She is a technical staff member in the AI Technology \& Systems group at MIT Lincoln Laboratory. Her research interests include human-centered AI, visual communication, and data science workflows.\end{IEEEbiography}

\begin{IEEEbiography}[{\includegraphics[width=1in,height=1.25in,clip,keepaspectratio]{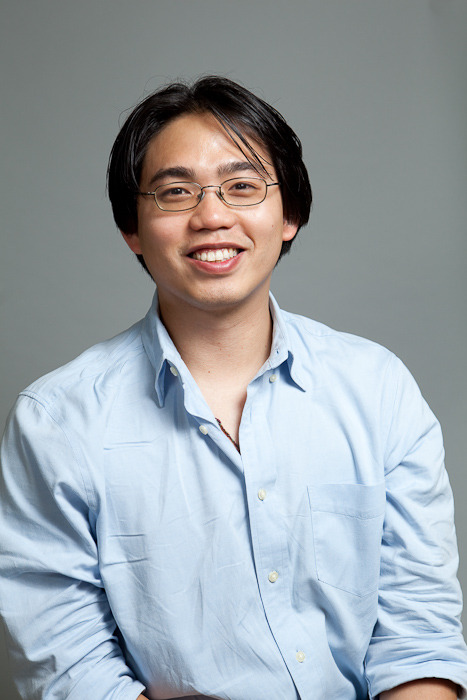}}]{Remco Chang} received his PhD in computer science from the University of North Carolina Charlotte. He is an associate professor in computer science with Tufts University. His research interests include visual analytics, information visualization, human computer interaction, and databases.\end{IEEEbiography}

\begin{IEEEbiography}[{\includegraphics[width=1in,height=1.25in,clip,keepaspectratio]{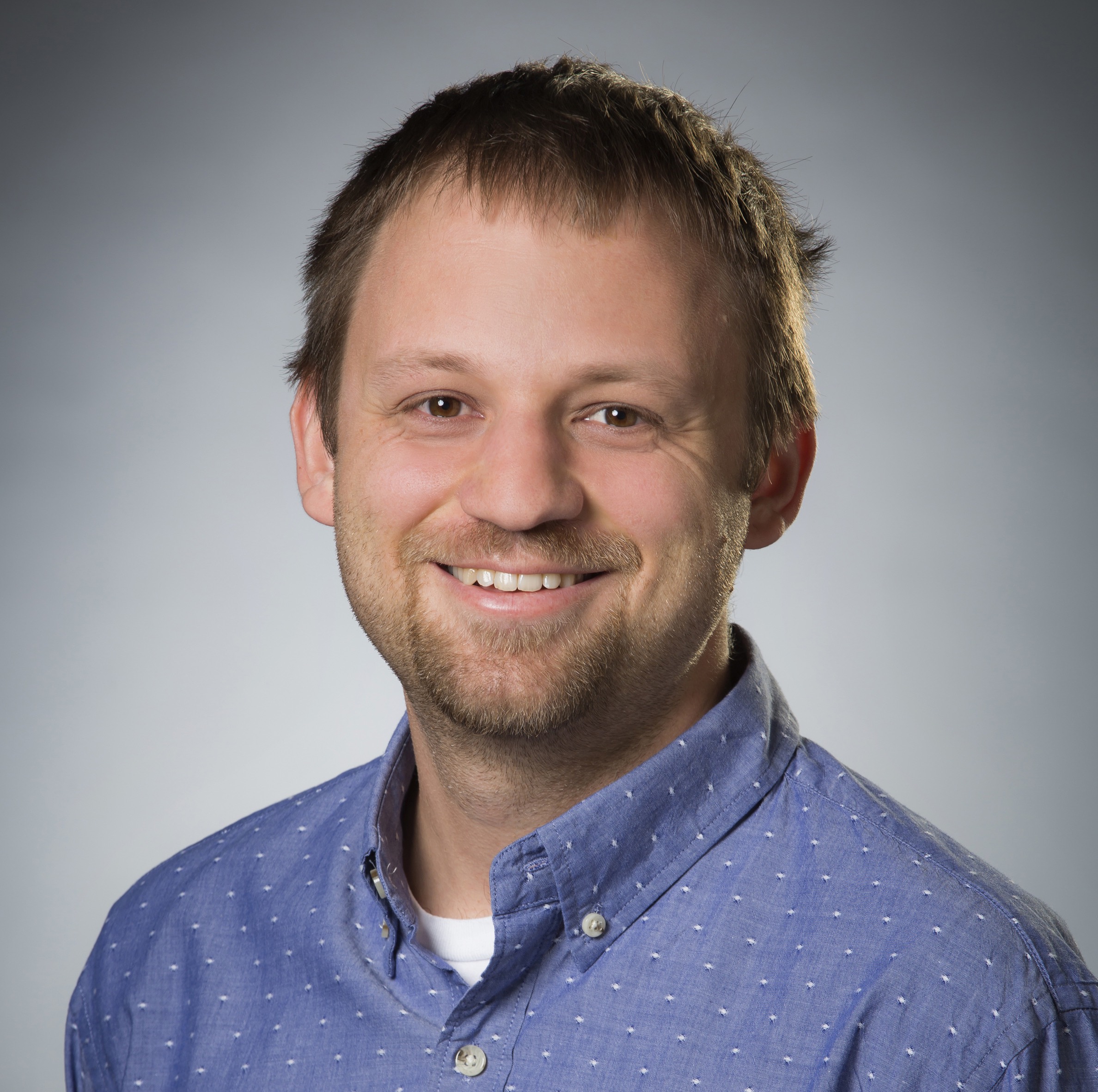}}]{Alex Endert} is an associate professor at the School of Interactive Computing, Georgia Tech. He directs the Visual Analytics Lab, which explores novel user interaction techniques for visual analytics, for domains including intelligence analysis, cyber security, manufacturing, decision making, and others.\end{IEEEbiography}

\end{document}